\documentclass[aip,apl,twocolumn,reprint]{revtex4-1}
\usepackage{graphicx}
\usepackage{dcolumn}
\usepackage{bm}
\usepackage[usenames,dvipsnames]{color} 
\usepackage{tabularx}
\usepackage{array}
\usepackage{amsmath}
\usepackage{pbox}
\usepackage{stmaryrd}  
\usepackage{floatrow}
\usepackage{sidecap}

\begin{document}

\title{Combinatorial split-ring and spiral \textcolor{black}{meta-resonator} for efficient magnon-photon coupling}

\author{Yuzan Xiong}
\affiliation{Department of Physics and Astronomy, University of North Carolina at Chapel Hill, Chapel Hill, NC 27599, USA}

\author{Andrew Christy}
\affiliation{Department of Physics and Astronomy, University of North Carolina at Chapel Hill, Chapel Hill, NC 27599, USA}
\affiliation{Department of Chemistry, University of North Carolina at Chapel Hill, Chapel Hill, NC 27599, USA}

\author{Yun Dong}
\affiliation{Department of Physics and Astronomy, University of North Carolina at Chapel Hill, Chapel Hill, NC 27599, USA}

\author{Andrew H. Comstock}
\affiliation{Department of Physics and Organic and Carbon Electronics Lab (ORaCEL), North Carolina State University, Raleigh,
NC 27695, USA}

\author{Dali Sun}
\affiliation{Department of Physics and Organic and Carbon Electronics Lab (ORaCEL), North Carolina State University, Raleigh,
NC 27695, USA}

\author{Yi Li}
\affiliation{Materials Science Division, Argonne National Laboratory, Argonne, IL 60439, USA}

\author{James F. Cahoon}
\affiliation{Department of Chemistry, University of North Carolina at Chapel Hill, Chapel Hill, NC 27599, USA}

\author{Binbin Yang}
\thanks{byang1@ncat.edu}
\affiliation{Department of Electrical and Computer Engineering
North Carolina A$\&$T State University, Greensboro, NC 27411, USA}

\author{Wei Zhang}
\thanks{zhwei@unc.edu}
\affiliation{Department of Physics and Astronomy, University of North Carolina at Chapel Hill, Chapel Hill, NC 27599, USA}

\date{\today}

\begin{abstract}

Developing \textcolor{black}{hybrid materials and structures} for electromagnetic wave engineering has been a promising route towards novel functionalities and tunabilities in many modern applications. Despite its established success in engineering optical light and terehertz waves, the implementation of \textcolor{black}{meta-resonators} operating at the microwave band is still emerging, especially those that allow for on-chip integration and size miniaturization, which has turned out crucial to developing hybrid quantum systems at the microwave band. In this work, we present a microwave \textcolor{black}{meta-resonator} consisting of split-ring and and spiral resonators, and implement it to the investigation of photon-magnon coupling for hybrid magnonic applications. We observe broadened bandwidth to the split ring modes augmented by the additional spiral resonator, and, by coupling the modes to a magnetic sample, the resultant photon-magnon coupling can be significantly enhanced to more than ten-fold. Our work suggests that combinatorial, \textcolor{black}{hybrid} microwave resonators may be a promising approach towards future development and implementation of photon-magnon coupling in hybrid magnonic systems.

\end{abstract}

\maketitle

\section{Introduction}

The emerging needs for quantum technologies require information propagation over long distances and with preserved coherence. This can be achieved by developing hybrid systems consisting of two subsystems operating in a strong-coupling regime \cite{kurizki_2015,nakamura_apex2019,hu_ssp2018,tqe_2021,jap_2021,prep_2022}. For this purpose, microwave resonators loaded by a low-damping ferrimagnet, e.g., Y$_3$Fe$_5$O$_{12}$ (YIG), implementing the strong photon-magnon coupling are often used \cite{serga_jphysd2010}. Coherent transfer of information can be subsequently achieved between electromagnetic (EM) carrier to various other excitations, such as spin oscillations, acoustic phonons, and qubits \cite{huebl_prl2013,flatte_prl2010,xufeng_sciadv2016,xufeng_prl2014,nakamura_science2015,an_prb2020,fukami_prxq2021}.  

\begin{figure*}[htb]
 \centering
 \includegraphics[width=5.9 in]{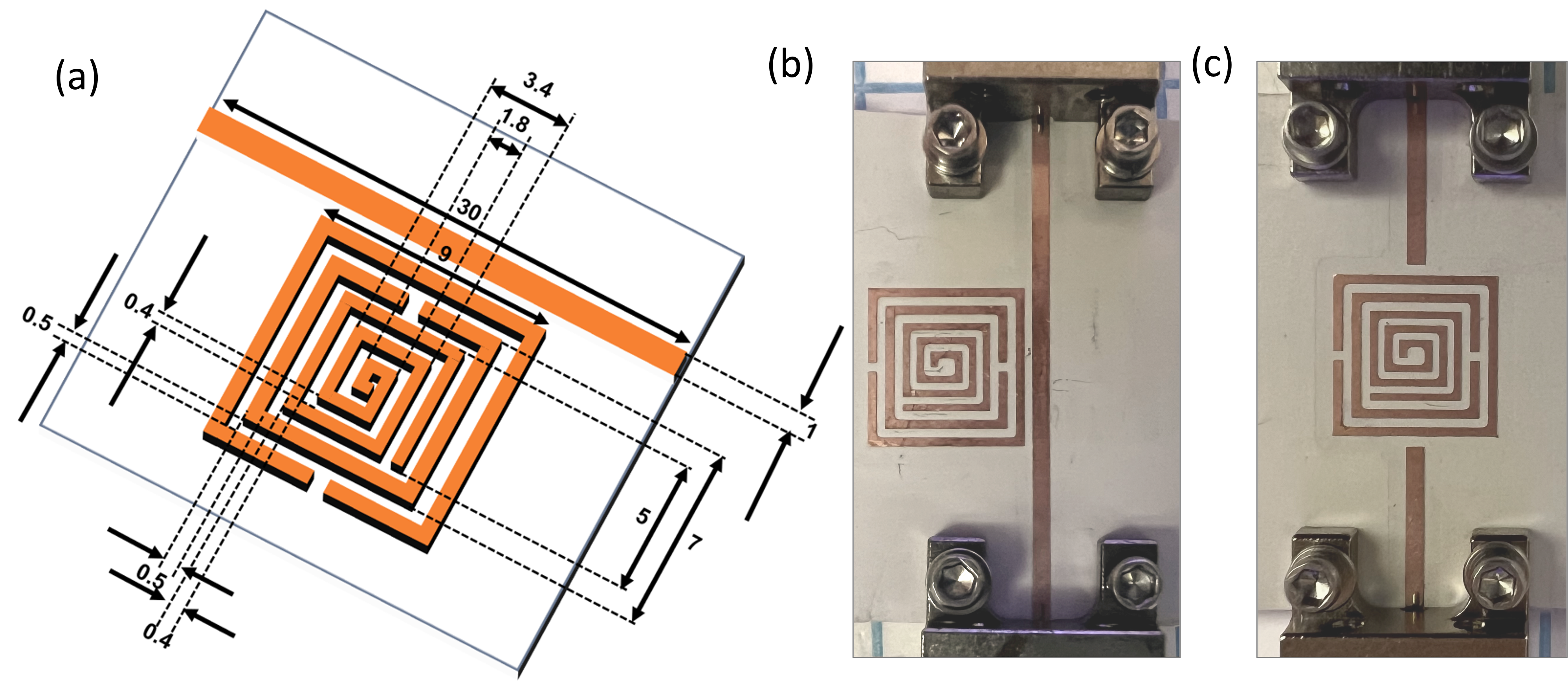}
 \caption{(a) Schematic illustration and dimensional specification of the s-DSRR \textcolor{black}{meta-}resonator. Unit is in mm. The \textcolor{black}{meta-}resonator is implemented to either (b) the transmission-line coupled (inductive) or (c) the edge-coupled (capacitive) MW feeding mechanism. }
 \label{fig1}
\end{figure*}

Among various strategies for improving the photon-magnon coupling strength, engineering the microwave (MW) photon mode, whether in 3-D cavities \cite{tobar_prap2014}, planar resonators \cite{bhoi_ssp2019,bhoi_srep2017,bhoi_jap2022}, or cryogenic platforms \cite{yili_prl2019,luqiao_prl2019,yili_prr2023,igor_sciadv2021}, remains an effective approach for realizing strongly coupled systems, as the coupling strength is directly proportional to geometrical considerations such as the filling factor of the magnon counterpart by the magnetic component, $h$, of the resonator EM field \cite{tobar_njp2019}. In particular, the design of a split-ring resonator (SRR) and its derivatives have been widely adopted in numerous photon-magnon systems, due to the various advantages including but not limited to subwavelength operation, negative permittivity and permeability, and broadband circuit integration  \cite{bhoi_prb2019,weiler_apl2016,chai_jphysd2019,inman_prap2022,vakula_srep2023}.

In general EM wave engineering, using \textcolor{black}{hybrid} structures allows strong localization and field enhancement which enable important novel functionalities with improved compatibility for various applications \cite{padilla_am2012}. The EM properties can be controlled by tailoring the geometric dimension and metallic structure, whose enlarged design parameter space allows to access and improve characteristics of the bandwidth, frequency selectivity, polarization and size miniaturization. Besides, by further confining the EM fields at the surface through localized modes, surface plasmonic \textcolor{black}{meta-resonators} are particularly promising for applications that require ultra-strong matter-field interactions in the near-fields \cite{cui_rmp2022,helin_ieee2022}. In this regard, great success of using \textcolor{black}{meta-}resonators has been demonstrated primarily in the terahertz (THz) EM engineering \cite{averitt_nature2006,schultz_prl2000}, however, still remains elusive for the photon-magnon systems in the MW band.

In this work, we present a combinatorial \textcolor{black}{double-split-ring resonator (DSRR)} incorporating a spiral-shaped, \textcolor{black}{meta-}resonator for hybrid magnonic applications. Compared to the single split-ring that has been dominantly used, the use of a double-split-ring spectrally extends the bandwidth to a lower frequency region, and more importantly, spatially re-distribute the magnetic field ($h-$component) towards the center of the resonator \cite{guo_oe2007}. The latter is particularly important for additional metallic constructions inside the center dielectric area. On the other hand, the spiral resonator has been frequently implemented in miniaturized and multi-resonant MW circuit applications such as wireless sensing, selective filters, and chipless RFID Tags \cite{swissroll,air_srep2022,vegni_ieee2007,lopes_ieee2020}. However, a generic spiral resonator does not couple effectively to an adjacent stripline and the amplitude usually decays rather rapidly as the distance to the MW transmission line increases. 

Here, we incorporate a spiral resonator in the center of the DSRR and construct a metamaterial MW resonator with structural symmetry breaking, termed as spiral(s)-DSRR, leveraging its multi-resonance characteristics and an enhanced magnetic coupling (to magnetic samples) due to the increased filling factor and the formation of localized, trapped microwave modes \cite{sova_ieee2022,an_prap2023,zheludev_prl2007}. 

We study the spatial properties of the electric- and magnetic-field distribution of the \textcolor{black}{meta-}resonator and investigate its application in the context of strong photon-magnon coupling with magnetic YIG samples. We show that additional MW photon modes emerge by adding a spiral resonator to the generic DSRR, and the resultant photon-magnon coupling can be significantly enhanced to more than ten-fold. The results are further corroborated by using microwave simulations. Our design of a combinatorial DSRR and spiral resonator may be treated as a general design protocol for the unit-cell of MW metamaterial future construction of photon-magnon coupled systems. 

\begin{figure*}[htb]
 \centering
 \includegraphics[width=7.5 in]{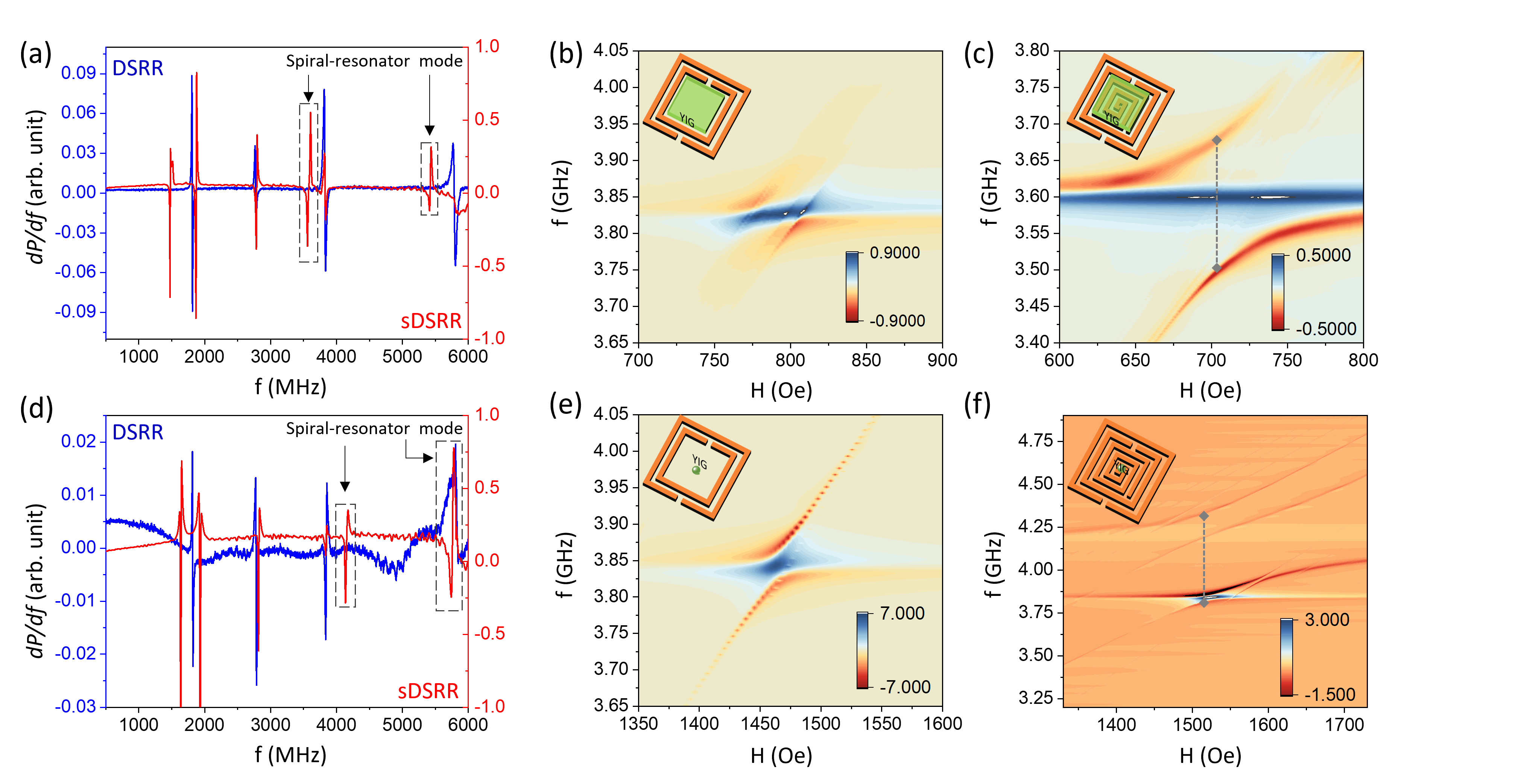}
 \caption{Results for the transmission-line coupled (inductive) resonator: (a) Absorption spectrum $dP/df$ of the DSRR and s-DSRR when a YIG planar film is loaded. Additional peaks emerge for the s-DSRR compared to the generic DSRR. 2D contour plot of the [$f,H$]-dispersion showing the photon-magnon coupling: (b) with the DSRR photon mode at 3.83 GHz, and (c) with the s-DSRR photon mode at 3.58 GHz. (d) Absorption spectrum $dP/df$ of the DSRR and s-DSRR when a YIG sphere is loaded. 2D contour plot of the [$f,H$]-dispersion showing the photon-magnon coupling: (e) with the DSRR photon mode at 3.85 GHz, and (f) with the s-DSRR photon mode at 4.16 GHz.  }
 \label{fig2}
\end{figure*}

\section{Experiment and results}

\subsection{Design and photon resonances of s-DSRR}

Figure \ref{fig1}(a) illustrates our key design of the combinatorial s-DSRR. The outer ring of the DSRR has a dimension of $9 \times 9$ mm$^2$, with a conductor width $w$ = 0.5 mm. the inner ring has a dimension of $7 \times 7$ mm$^2$, and is deployed at a distance of 0.5 mm from the outer ring. The split portion of the inner ring has a width of 0.5 mm. The \textcolor{black}{meta-}resonator consists of the DSRR with the same dimension mentioned above and an additional spiral loop in the middle of the resonator dielectric region. The distance of the spiral to the inner ring of the DSRR is 0.4 mm. 

We implement the \textcolor{black}{meta-}resonator to either a transmission-line, Fig. \ref{fig1}(b), or an edge-coupled microstrip, Fig. \ref{fig1}(c). In the former, the distance between the outer ring to the feeding transmission line, which is 1 mm wide, is 0.45 mm. In the latter, the gap between the microstrip and the outer ring is 0.6 mm. All resonators and the microstrips (transmission or edge-coupled) that feeds them were fabricated on the Rogers TMM Laminates dielectric substrate (Rogers Corp.) with dielectric properties ($\epsilon_r$) of 9.8, tangent loss (tan$\delta$) of 0.0020, substrate thickness of $T_s$ = 1.27 mm, and a copper layer thickness of $T_{Cu}$ = 35 $\mu$m.

The split-ring structure introduces the inductive and capacitive responses to the time-varying electric and magnetic fields from the transmission line or the microstrip. The use of DSRR spatially re-distributes the magnetic field ($h-$component) towards the center of the resonator (in the dielectric) and with higher uniformity, as opposed to localize on the resonator side near the gap to the MW feed. The spiral loop inside the DSRR is then introduced in the middle. Such a combination enhances the magnetic coupling between the magnetic sample and the subwavelength-based \textcolor{black}{meta-resonator} structure, and gives rise to improved polarization characteristic which further enhances the negative electric and magnetic susceptibility. The metallic layer of copper exhibits the inductive response, where as the dielectric substrate and the space between the metallic arms, are responsible for the capacitive response. For the edge-coupled stripline in Fig. \ref{fig1}(c), the gap between the feedline and the resonator acts as a parallel plate with additional capacitance to further enhance the qualify(Q)-factor but at a cost of higher insertion loss \cite{edge_coupled}. Inside the s-DSRR region, the inductive properties of the metallic layer dominate over the capacitive value of dielectric substrate, which is the key to an enhanced photon-magnon coupling.

\begin{figure*}[htb]
 \centering
 \includegraphics[width=5.5 in]{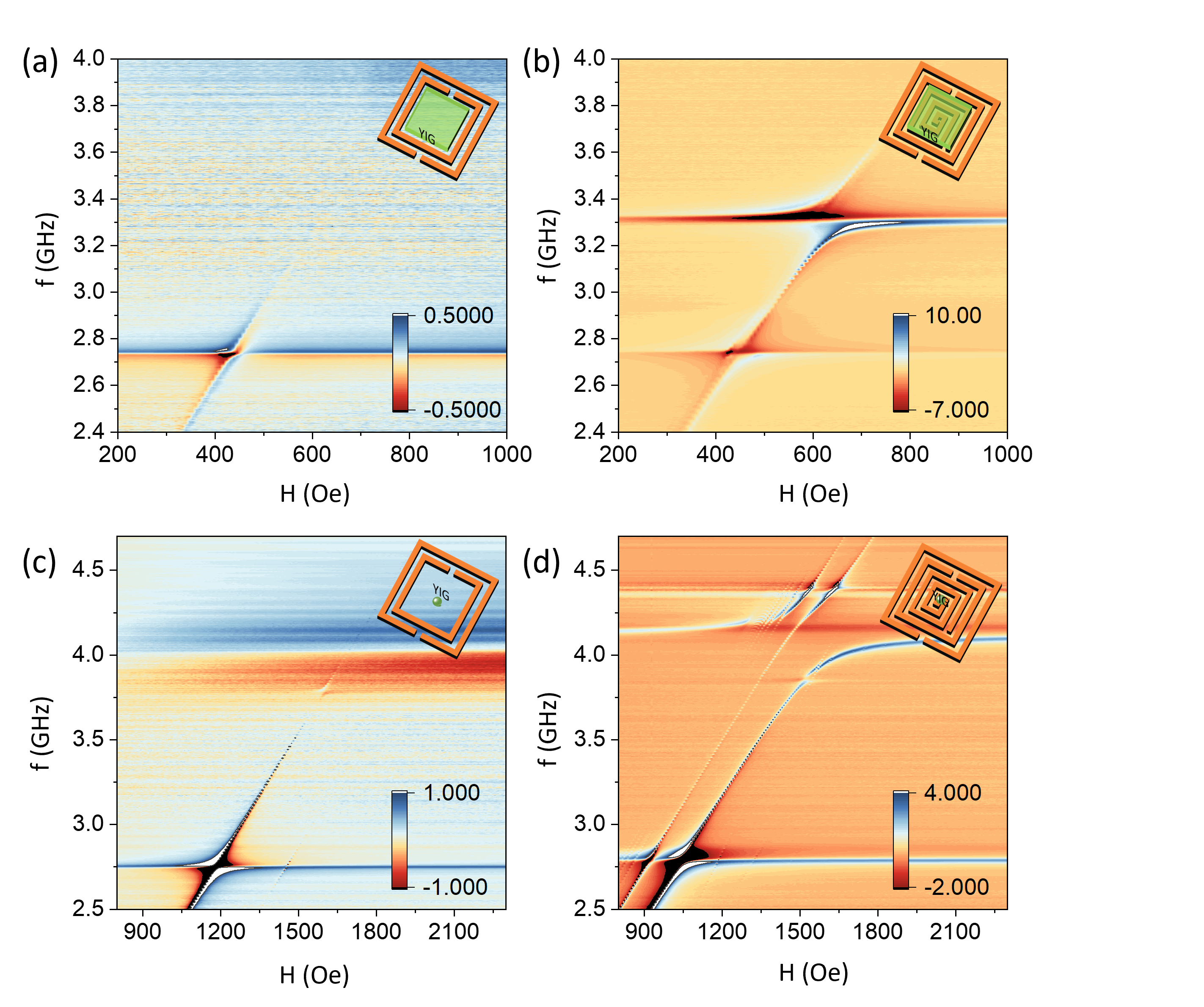}
 \caption{Results for the edge-coupled (capacitive) resonator: 2D contour plot of the [$f,H$]-dispersion showing the photon-magnon coupling. For YIG film: (a) with the DSRR resonator (trivial coupling), and (b) with the s-DSRR photon mode at 3.31 GHz. For YIG sphere: (c) with the DSRR photon mode (trivial coupling), and (d) with the s-DSRR photon mode at 4.16 GHz.}
 \label{fig3}
\end{figure*}

\begin{figure*}[htb]
 \centering
 \includegraphics[width=6.9 in]{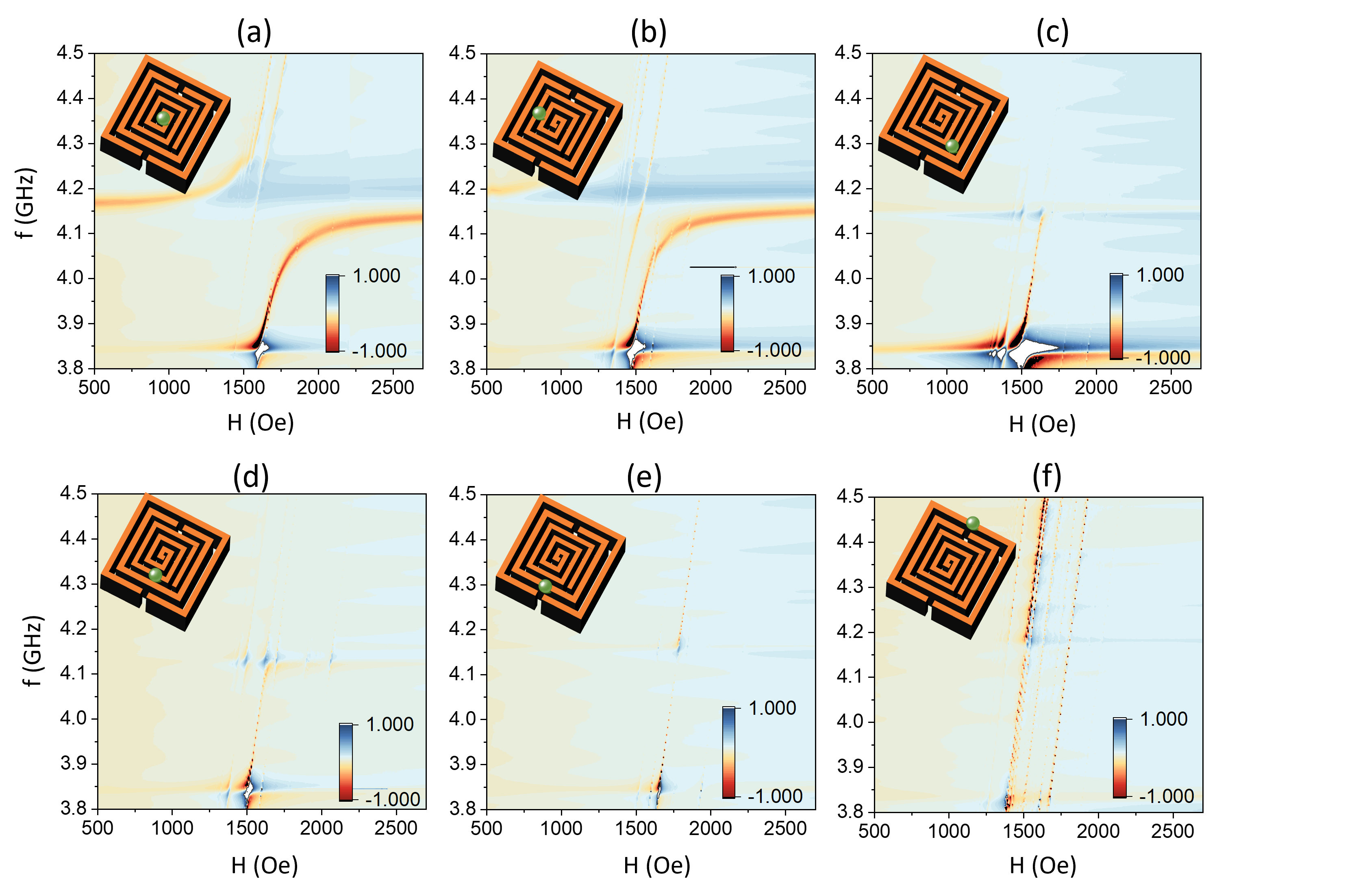}
 \caption{2D contour plot of the [$f,H$]-dispersion showing the photon-magnon coupling between the YIG sphere and the s-DSRR photon modes at different representative locations: (a) center of spiral, (b) inner arm of spiral, (c) tail of spiral, against DSRR inner corner, (d) inner arm of DSRR, (e) outer arm of DSRR, away from stripline, and (f) outer arm of DSRR, close to stripline. }
 \label{fig4}
\end{figure*}

\subsection{Photon-magnon coupling of s-DSRR}
\label{section:coupling}
We first focus on the \textcolor{black}{meta-}resonator coupled to the transmission line, Fig. \ref{fig1}(b), and compare the spectra characteristics of the s-DSRR and the pristine DSRR, measured by using a vector-network analyzer (VNA) up to 6 GHz with the application of an external magnetic field, $H$, perpendicular to the transmission line, to satisfy the ferromagnetic resonance condition. The resonance peaks (photon modes) are more clearly identified by the derivative absorption $dP/df$ than the conventional S$_{21}$. 

We investigate the photon-magnon coupling of our s-DSRR to magnetic YIG samples. Both planar YIG film (cut into the size of the spiral resonator) and YIG spheres are used. The thickness of the YIG film is 2.0 $\mu$m on a 0.5 mm thick gadolinium gallium garnet (GGG) substrate. The nominal diameter of the YIG sphere is 1.0 mm. After loading the YIG sample (either the planar film or the sphere), slight shifts of the photon modes can be observed due to the change of impedance. In addition, new modes emerge which are unique to the s-DSRR compared to the original DSRR. Figure \ref{fig2}(a) summarizes the absorption spectra ($dP/df$) of the DSRR and the s-DSRR with a planar YIG film load. The DSRR has four pronounced photon modes, at 1.81, 2.77, 3.83, 5.79 GHz. The spiral resonator introduces additional modes at 1.48, 3.58, 5.42 GHz. To quantify and compare the MW photon-magnon coupling, the coupling strength, $g$, is typically used, which is directly obtainable from the experimentally (or numerically) acquired transmission spectra characterized by an anti-crossing gap \cite{huebl_prl2013,flatte_prl2010,xufeng_sciadv2016,xufeng_prl2014,nakamura_science2015,an_prb2020,fukami_prxq2021}. 

The coupling of all four DSRR modes to the YIG planar film were rather insignificant. For example, Fig. \ref{fig2}(b) shows the [$f,H$] dispersion near a DSRR photon mode around 3.83 GHz. Only intermediate coupling strength is evident without the observation of a strong anti-crossing feature. However, another adjacent mode at 3.58 GHz that was caused by the spiral resonator exhibits a strong coupling with the YIG film, with a pronounced anti-crossing feature, Fig. \ref{fig2}(c). The coupling strength is estimated as $g/2\pi = 92$ MHz.  

The same effect is also observed for using the YIG spheres. Likewise, Fig. \ref{fig2}(d) compares the DSRR and s-DSRR spectra with a YIG sphere load. Additional spiral resonator modes (at 1.64, 1.93, 4.16, 5.76 GHz) beyond the DSRR modes (at 1.82, 2.78, 3.85, 5.82 GHz) can be observed. The YIG sphere generally couples stronger than the planar YIG film, for example, as shown in Fig. \ref{fig2}(e), where a nontrivial anti-crossing feature corresponds to $g/2\pi = 25$ MHz is observed. However, when looking at the nearby spiral resonator mode at 4.16 GHz, the coupling strength increased by more than tenfold, as shown in Fig. \ref{fig2}(f), with a $g/2\pi = 265$ MHz. Notably, the previous anticrossing feature at around 3.85 GHz due to coupling to the DSRR mode is still seen, whose coupling strength remains similar in magnitude to the case in Fig. \ref{fig2}(e). 

Similar observation was also found for the other edge-coupled resonator, see Fig. \ref{fig3}. Such a configuration aims to increase the Q-factor via an enhanced capacitive coupling, but at a cost of additional microwave loss. Indeed, As shown in Fig. \ref{fig3}(a) and (c), the coupling using the generic DSRR photon mode at $\sim$ 2.75 GHz is weak for both planar and spherical YIG loading. However, strong coupling emerges with the additional spiral structure at the spiral resonator modes, and a coupling strength of $g/2\pi =$105 and 260 MHz were found for the YIG film, Fig. \ref{fig3}(b), and sphere, Fig. \ref{fig3}(d), respectively.

By using the YIG sphere whose size is considerably smaller (diameter $\sim$ 1.0 mm) than the dimension of the \textcolor{black}{meta-}resonator ($9 \times 9$ mm$^2$ with metal/gap width $\sim$ 0.5 mm), we can also study the spatial properties of the photon-magnon coupling at different locations of interest. Figure \ref{fig4} summarizes the photon-magnon coupling with the spiral resonator mode at 4.16 GHz by attaching the YIG sphere at representative locations. Overall, the coupling strength reduces as the YIG sphere is moved away from the center of the spiral resonator. Following the strongest coupling at the center (Fig. \ref{fig4}(a)), slightly reduced coupling is observed when the YIG is placed off the center, at the inner spiral arm, see Fig. \ref{fig4}(b). The coupling rapidly decreases when the sample if further off-centered, for example, at the outer spiral arm, see Fig. \ref{fig4}(c), when the sphere is placed at the tail-end of the spiral and facing against the corner of the inner-SRR loop. On the other hand, by the above movement, the photon-magnon coupling with the DSRR mode at 3.85 GHz becomes progressively stronger, due to the closer proximity to the DSRR metallic loop that features a dominant inductive coupling, although the coupling strength is still one order of magnitude smaller than spiral resonator mode. Last but not least, the photon-magnon coupling with both modes (s-DSRR: 4.16 GHz and DSRR: 3.85 GHz) is weak at the DSRR arm locations, for example, at the inner arm, Fig. \ref{fig4}(d), and at the outer arm, whether (the YIG sphere) is away, Fig. \ref{fig4}(e), or near, Fig. \ref{fig4}(f), the transmission line. 

Finally, similar conclusion can be drawn from the other higher spiral resonator mode, at 5.42 and 5.76 GHz for YIG film and sphere loads, respectively, in Fig. \ref{fig2}(a) and (d). Enhanced photon-magnon coupling with notable anti-crossing gaps were found at the s-DSRR modes, as compared to the generic DSRR modes.

\section{Simulation and modeling}

\begin{figure*}[htb]
 \centering
 \includegraphics[width=5 in]{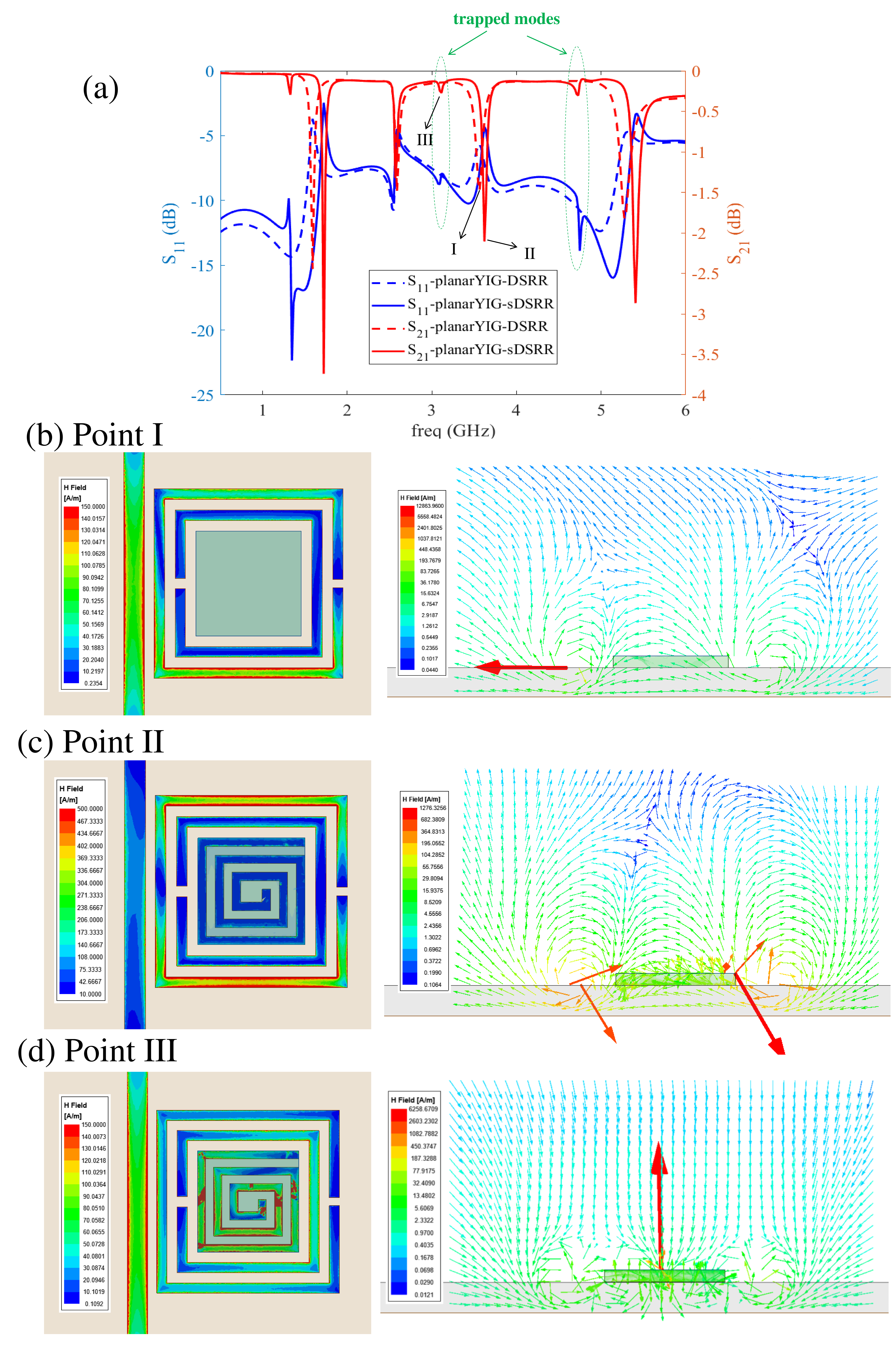}
 \caption{(a) the simulated reflection (blue) and transmission (red) spectrum of the planar YIG loaded DSRR (dashed lines) and s-DSRR (solid lines) configurations; and the magnetic field distribution on the resonator and in the near field region for (b) point I (regular resonance point for DSRR at 3.564 GHz), (c) point II (regular resonance for s-DSRR at 3.620 GHz) and (d) point III (trapped mode for s-DSRR at 3.11 GHz).}
 \label{fig5}
\end{figure*}

\begin{figure*}[htb]
 \centering
 \includegraphics[width=5 in]{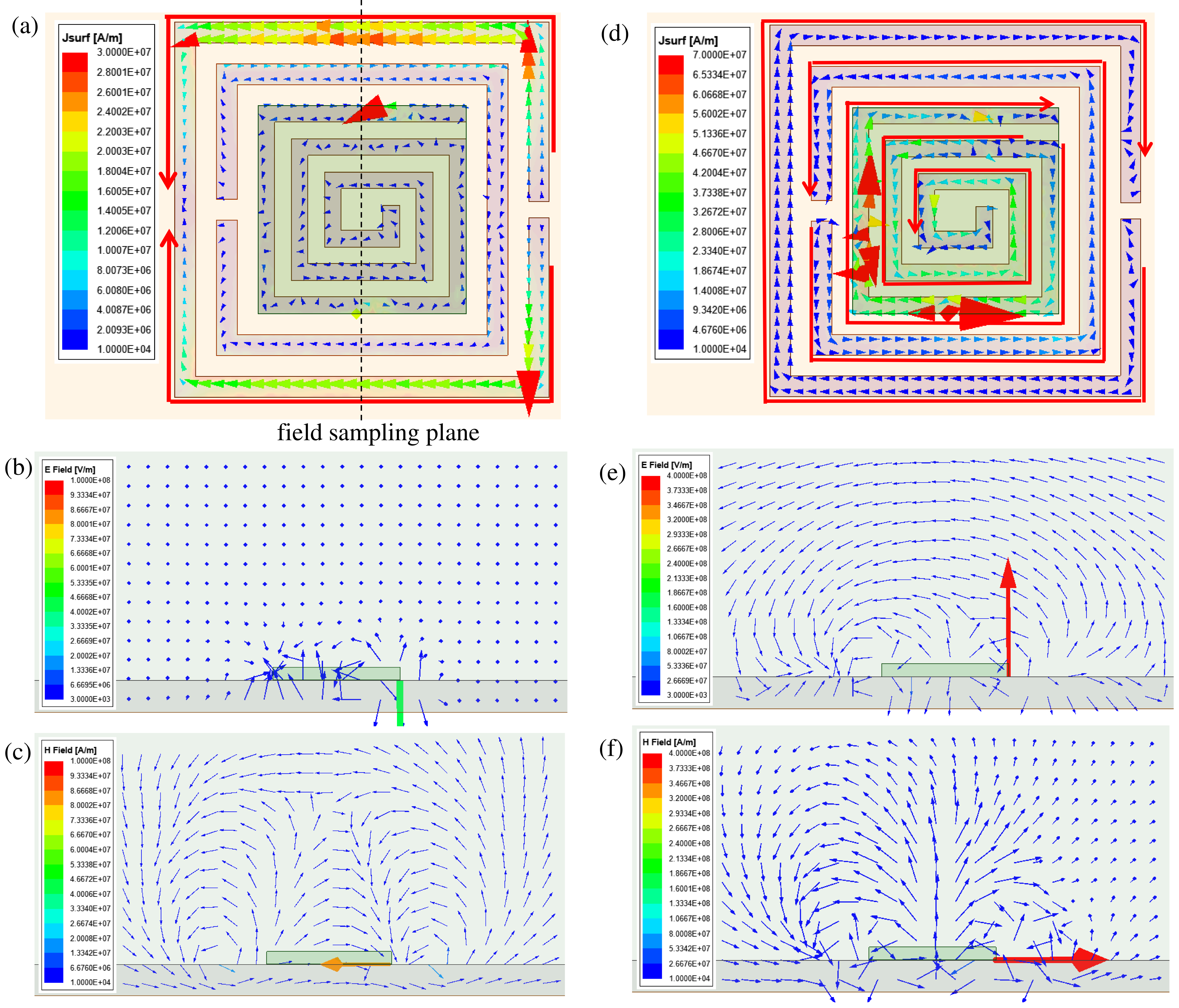}
\caption{ \textcolor{black}{The eigenmode surface current (a), eigen E field (b) and eigen H field (c) for the resonance at 3.62 GHz on s-DSRR; the eigenmode surface current (d), eigen E field (e) and eigen H field (f) of the spiral-induced trapped mode for s-DSRR at 3.11 GHz. Unit ``stored energy" in each mode is used for normalization.}
 }
 \label{fig5_2}
\end{figure*}

To gain an in-depth understanding of the enhanced coupling, we performed simulation and modeling of the YIG-loaded DSRR and s-DSRR using the High Frequency Structure Simulator (HFSS) from ANSYS. \textcolor{black}{The two-port microstrip line structure is used to excite the meta-resonators.} Both the planar YIG and the spherical YIG are investigated, with the dimensions of the YIG components given in section \ref{section:coupling}. Specifically, in HFSS, the permeability of the YIG component is modeled with the Polder's tensor \cite{polder_tensor} with the following material parameters: $M_s=1750$ G, $\Delta H=5$ Oe and lande $g-$factor of 2. A uniform magnetic biasing field $H_0=0$ Oe is applied horizontally. The dielectric permittivity and the loss tangent of the YIG components are $\epsilon_r=14$ and $\tan \delta=0.0001$ respectively. The GGG substrate has the permittivity of $\epsilon_r=11.99$ and the loss tangent $\tan \delta=0.0052$. In both the planar the spherical YIG cases, the s-DSRR is found to generate more resonant modes and significant frequency shift due to strong coupling. 



To elucidate the nature of the extra resonances, we here take the planar YIG as an example and investigate the EM field distribution. As shown in Figure \ref{fig5}, three extra sharp resonances are observed in the s-DSRR case shown in solid lines, in contrast to the DSRR case in dashed lines. This agrees reasonably close with the measurement result in Figure \ref{fig2} (a). Comparing the two sets of data in Figure \ref{fig5}, it is evidenced that the extra resonances noticed in the s-DSRR structure are created due to the spiral structure. \textcolor{black}{As a further investigation, an eigenmode simulation of the s-DSRR resonator is conducted in Ansys HFSS. Similar resonances as reported in Figure \ref{fig5}(a) are observed in the eigenmode simulation.} \textcolor{black}{In addition, Fig. \ref{fig5_2}(a) and (d) show the surface current of the two modes at 3.62 GHz and 3.11 GHz respectively. While the resonance at 3.62 GHz is dominated by the split-ring structure with primarily in-phase currents, the resonance at 3.11 GHz has dominant currents distributed over the spiral region with primarily anti-phase currents on all neighboring conductors. The anti-phase currents result in very weak coupling to radiation in free space, a key feature of the trapped modes \cite{trapped_modes}.} 

Next, we compare the simulated magnetic field distribution for the DSRR and s-DSRR with the planar YIG load. As an example, we focus on the spiral resonator modes with pronounced photon-magnon coupling, \textcolor{black}{at 3.58 GHz in measurement (3.11 GHz in simulation) for the planar YIG. Figure \ref{fig5}(b) and (c) plot the dynamic magnetic fields on the metallic structure and in dielectric near field region} of the regular (bright) modes, for the YIG-film-loaded DSRR and s-DSRR, respectively. For the DSRR, the strongest magnetic fields reside around the metallic arms of the split-structures. The magnetic fields inside the central dielectric area are overall weak, and are primarily polarized in the resonator plane. For the s-DSRR, on the other hand, the magnetic fields are significantly enhanced inside the central dielectric region due to the densely confined surface currents on and near the spiral metallic arms. The vector fields are strongly polarized perpendicular to the plane, which enhances the applied excitation along both the horizontal and vertical plane of the \textcolor{black}{meta-}resonator. \textcolor{black}{As a validation check, the eigen E and H fields for the resonances at 3.62 GHz and 3.11 GHz are also provided in Figure \ref{fig5_2}(b)-(c) and (e)-(f) respectively. Similar magnetic field enhancement is observed for the trapped mode at 3.11 GHz.}

In addition, Figure \ref{fig5}(d) plots the field distribution for the trapped mode at 3.11 GHz, which is unique to the s-DSRR. An even stronger field concentration at the spiral resonator region and near the surface of the dielectric can be evidenced. The inner spiral arms and the spiral center also show stronger fields than the outer spiral arms. This 'radial-like decay' is in agreement with the general features of such spiral resonators, due to the radial decay of the structural asymmetry as the spiral arms become longer \cite{air_srep2022, vegni_ieee2007, lopes_ieee2020}. Such a characteristic is also consistent with the earlier position-dependent measurements in Fig. \ref{fig4}, where the photon-magnon coupling quickly weakens as the YIG sphere is gradually moved away from the spiral structure.

\section{Analysis and Discussion}

Phenomenologically, the value of coupling strength $g$ is directly proportional to the form factor $\eta$ of the magnetic sample by the magnetic component of the resonator's EM field, $h$, that is perpendicular to the external bias field, $H$, and the coupling strength can be written \cite{tobar_njp2019}:    
\begin{equation}
    g=\frac{\gamma}{2} \eta \sqrt{\frac{\mu_0 S \hbar \omega_p}{V_m}},
\label{Eq:g}
\end{equation}
where $\gamma$ is the gyromagnetic ratio, $S$ is the total spin number, whose value is proportional to the magnetic moment of the magnetic sample and the number of spins ($N_s$), $\hbar$ is the reduced Plank constant, $\mu_0$ is the vacuum permeability, and $\omega_p$ is the photon mode frequency. The spin concentration ($n_s$) then relates to the total spins of the sample and the magnetic sample volume, by $n_s = N_s/V_m$. When the form factor $\eta=0$, none of the resonator's $h$-field is perpendicular to the external bias field, whilst when $\eta=1$, all of the $h$-field is perpendicular to the sample. 

Notably, the role of $\eta$ has been traditionally overlooked, and an estimation of the relationship: $g \propto \sqrt{N_s}$ is often claimed, hinting further: $g \propto \sqrt{V_m}$. However, increasing the volume of the sample also changes the $\eta$, therefore affecting the coupling strength. Such an effect can be corrected by taking into account the geometrical configurations like the magnetic filling factor, $\zeta_m$, of cavity field contained in the sample by $g=\omega_c \sqrt{\chi_\textrm{eff} \zeta_m}$, where $\chi_\textrm{eff}$ is an effective susceptibility determined by material properties and the overlap of the specific cavity and magnon modes \cite{tobar_prap2014}. Therefore, the magnetic filling factor becomes the relevant figure of merit in maximizing the coupling strength. 


In our \textcolor{black}{hybrid} design, the DSRR component induces a negative magnetic susceptibility for applied transverse EM waves, and its combination with a spiral resonator improves the electric and magnetic polarization and therefore further enhances negative electric and magnetic susceptibility. The mode volume of such dark mode is constricted due to its near-field confinement, which enables a strong spatial overlapping between the photon and magnon modes. Thus, the magnetic field distribution across the spiral structure increases the magnetic coupling within the spiral loop area and therefore enhances the magnetic filling factor of the sample \cite{tobar_njp2019}. 

Theoretically, The interaction \cite{hu_ssp2018} between the individual photonic components of this system (either the spiral or the DSRR) and the YIG sample can be modeled by two 
coupled classical oscillators \(O_1\) and \(O_2\). Their equations of motion can be written as \cite{tay_jmmm2018}: 
\begin{equation} \label{eq:1}
\ddot{x_1} + \omega_c^2 x_1 + 2\beta\omega_c\dot{x_1} - gx_2 = fe^{-i\omega t} 
\end{equation}
\begin{equation} \label{eq:2}
\ddot{x_2} + \omega_m^2 x_2 + 2\alpha\omega_c\dot{x_2} - gx_1 = 0 
\end{equation}

\noindent where \(O_1\) represents the s-DSRR and is driven by a plunger with a constant frequency \(\omega\). \(O_1\) is connected to the plunger by a spring with resonant frequency \(\omega_c\). The plunger represents the microwave transmission line and applies a driving force to the spring. \(\beta\) is the damping due to a viscous force, \(\alpha\) is the Gilbert damping factor, and \(g\) is the coupling strength between the two oscillators. \(O_2\) is connected to a fixed wall by another spring with resonant frequency \(\omega_m\). To explore the effect of the coupling strength \(g\) and by extension, the magnetic filling factor \(\zeta_m\), the dispersion of this system must be calculated. \(x_1\) and \(x_2\) can be expressed as plane waves \((x_1, x_2)\) = \((A_1, A_2)e^{-i\omega t}\). Using this representation, Equations (2) and (3) can be written in matrix form as:  
\begin{equation}
\begin{pmatrix} 
\omega^2 - \omega_p^2 + 2i\beta\omega_c\omega & g\\ g & \omega^2 - \omega_m^2 + 2i\alpha\omega_c\omega 
\end{pmatrix}
\begin{pmatrix}
A_1\\A_2
\end{pmatrix}
= \begin{pmatrix}
-f\\0
\end{pmatrix}
\end{equation}

\begin{figure*}[htb]
 \centering
 \includegraphics[width=7 in]{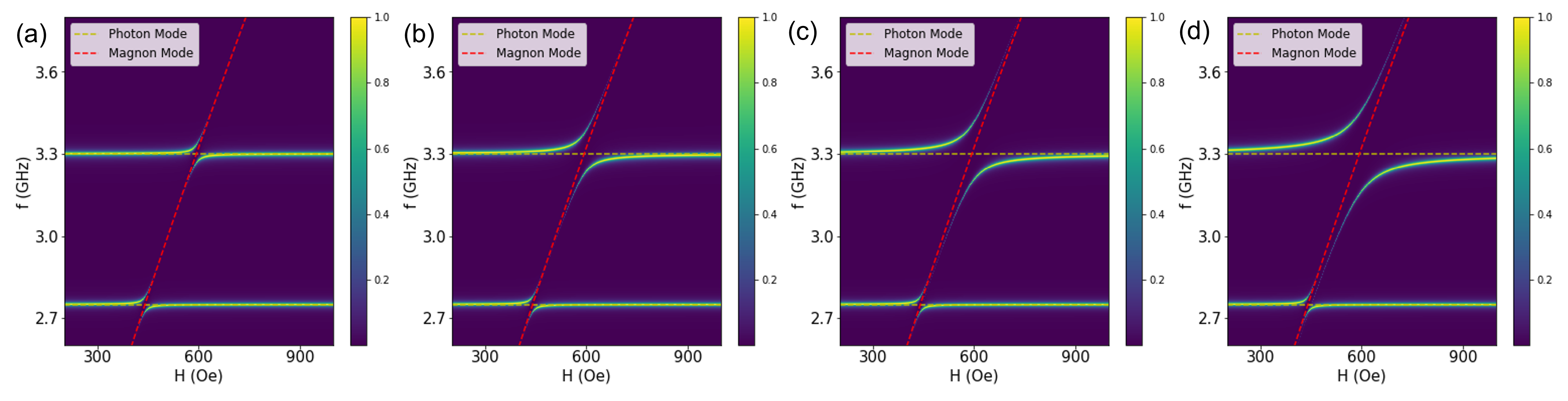}
 \caption{Variation of the quantity \(\tau\) (a) \(\tau\) = 1  (b) \(\tau\) = 5 (c) \(\tau\) = 10 (d) \(\tau\) = 20. The magnon mode for a planar YIG film was calculated by using the Kittel formula with a gyromagnetic ratio (\(\gamma\)) of 28 GHz/T ($g$-factor of 2) and a magnetization, $M$, of 0.175 T, \textcolor{black}{and a uniaxial anisotropy energy, \(H_A\), of 33.7 mT}. }
 \label{fig7} 
\end{figure*}

The dispersion can be found by calculating the poles of the determinant of the interaction matrix \textbf{M}. Since the coupling between the photon and magnon modes is what is being investigated, the region near where \(\omega_c = \omega_m\) is all that is relevant. In this region, the detuning \(\Delta\omega<< \omega_c + \omega_m\), so a rotating wave approximation can be made \cite{rotatingwaveapprox}. This simplifies the determinant to a form that can be easily solved. The dispersion, incorporating the magnetic filling factor \(\zeta_m\) \cite{tobar_prap2014} discussed earlier, simplifies to:

\begin{widetext}
\begin{equation}
\omega_{\pm} = \frac{1}{2} \left[\omega_m - \omega_c + i\omega_c(\alpha - \beta) \pm \sqrt{(\omega_m - \omega_c + i\omega_c(\alpha - \beta))^2 + 4\omega_c^2\chi_{eff}\zeta_m}\right]
\end{equation}
\end{widetext}

To evaluate the effect of the magnetic filling factor on the dispersion, the difference between the two solutions of the dispersion at the resonance condition (\(\omega_c = \omega_m\)) must be calculated: 
\begin{equation}
\omega_{gap} = (\omega_+ - \omega_-)\rvert_{\omega_c = \omega_m}
\end{equation}
\begin{equation}
\omega_{gap} = \omega_c\sqrt{4\chi_{eff}\zeta_m - (\alpha - \beta)^2}
\end{equation}
Experimentally determined values of these constants show that \(g^2 >> \alpha - \beta\), so \(\omega_{gap} = 2\omega_{c}\sqrt{\chi_{eff}\zeta_m}\).
Since \(O_1\) represents the s-DSRR, the microwave transmission \(S_{21}\) is proportional to \(A_1\). Therefore, solving Eq. 4 for \(A_1\) allows us to construct the \(|S_{21}|^2\) spectrum. In this system, \(\omega_c\) is the mode determined by the structure of the resonator and has no magnetic field dependence. However, \(\omega_m\) has a field dependence determined by the shape of the YIG sample. For the samples investigated above, the field dependences are represented by the Kittel equations for a planar magnetic film, \textcolor{black}{ $\omega_m = \gamma \sqrt{(H + H_A)(H + 4\pi M + H_A)}$}, where \(M\) is the magnetization of the sample and \(\gamma\) is the gyromagnetic ratio, \textcolor{black}{and \(H_A\) is the uniaxial anisotropy energy}. Using Fig. \ref{fig3}(b) as an example, modeling the additional spiral component mode at \textcolor{black}{3.31} GHz with a YIG \textcolor{black}{planar film} with \(\gamma\) = 28 GHz/T, the normalized \(|S_{21}|^2\) spectrum can be constructed. Since \(\chi_{eff}\) is determined by material properties and its change is negligible with the addition of the spiral component, the increase in coupling strength \(g\) of the magnon to the spiral photon modes can be attributed to the associated increase in \(\zeta_m\) \cite{tobar_prap2014,ultrahighcoop}.

The basic interaction of a single photon mode with a single magnon mode can pave the way for analysis of more complicated interactions. According to experiments in Fig. \ref{fig3}(b), we found much stronger coupling of the YIG to the spiral mode at 3.3 GHz relative to the DSRR mode at 2.75 GHz. Since the spiral component alone would not couple strongly to the transmission line, the interaction can be modeled as an initial coupling between the DSRR mode and the magnon mode, followed by a second coupling between the spiral mode and the resultant cavity-magnon-polariton mode. Both interactions could be modeled simultaneously using a 3 x 3 interaction matrix, but this complicates the dispersion calculation immensely. The consecutive matrices method outlined above allows for the use of two separate 2 x 2 interaction matrices, and is also more physically consistent with this process. The first is identical to Eq.4, and the second replaces the magnon mode with the DSRR polariton mode dispersion. Since the coupling occurs in two steps, we can define a \(\zeta_m^{DSRR}\), which is the magnetic filling factor of just the YIG/DSRR system, and \(\zeta_m^{spiral}\), which is the filling factor of the combined YIG/DSRR/spiral system. Here we define the quantity \(\tau\), which is the ratio of \(\zeta_m^{spiral}\) to \(\zeta_m^{DSRR}\). 

The effect of changing \(\tau\) on the \(|S_{21}|^2\) spectrum is shown in Figure \ref{fig7}. At \(\tau\) = 1, the coupling strengths are identical. As \(\tau\) increases, the coupling strength of the YIG to the spiral photon mode also increases, approaching the strength shown in Fig.\ref{fig3}(b). This shows that the increase in coupling strength of the magnon to the additional spiral modes relative to the DSRR modes is a result of the increase in magnetic filling factor described in our coupled harmonic oscillator model.  

In summary, we fabricated and characterized a combinatorial double-split-ring and spiral \textcolor{black}{meta-}resonator, and demonstrated its application for photon-magnon coupling in hybrid magnonics. Compared to the traditional single split-ring, the use of a double-split-ring spectrally extends the bandwidth to a lower frequency region and spatially re-distribute the magnetic field towards the center of the resonator. In addition, the double-split-ring allows to further couple it to an additional metallic spiral resonator. The spiral resonator photon modes can be effectively excited owing to its strong coupling to the double-split-ring, which further gives rise to multi-resonance characteristics as well as an enhanced magnetic coupling (to magnetic samples) due to the increased filling factor and the formation of localized surface dark modes. Our results imply that the dark modes arising from \textcolor{black}{meta-}resonators hold a great promise towards strong photon-magnon coupling at room temperature, and have the potential to be miniaturized and integrated with auxiliary optical systems \cite{o1,o2,o3}. 

\section*{Acknowledgment}
The experimental work at UNC-CH was supported by the U.S. National Science Foundation (NSF) under Grant No. ECCS-2246254. The computational work at NCAT was supported by the U.S. NSF under Grant No. ECCS-2138741.  Y. L. acknowledges support by the U.S. DOE, Office of Science, Basic Energy Sciences, Materials Sciences and Engineering Division under Contract No. DE-SC0022060. D.S. acknowledges support by the U.S. DOE, Office of Science, Basic Energy Sciences, under Contract No. DE-SC0020992.

\end{document}